\documentclass[ ]{revtex4}
\usepackage{hyperref}
\usepackage{amsmath}

\usepackage{amsfonts}

\begin{document}
\title{\bf \Large  Ward and Nielsen Identities  for   ABJM Theory in ${\cal N}=1$ Superspace }

 \author { Sudhaker Upadhyay }
 \email{ sudhakerupadhyay@gmail.com}
\affiliation { Centre for Theoretical Studies,\\
Indian Institute of Technology Kharagpur,  Kharagpur-721302,  India}

\begin{abstract}
The structures and the associated gauge algebra of  ABJM theory in ${\cal N}=1$ superspace 
are reviewed. We derive the Ward identities of the theory in the class of Lorentz-type gauges at quantum level to justify the renormalizability of the model. We compute the Nielsen identities for the two-point functions of the theory with the help of enlarged BRST transformation. The identities are 
derived in ABJM theory to ensure  the gauge independence of the physical poles of the Green's functions. 
\end{abstract}

\maketitle
 

\section{Introduction}
 Aharony, Bergman, Jafferis and Maldacena (ABJM) \cite{ABJ}
 have been found a breakthrough in understanding M2-branes in M-theory that the worldvolume theory of $N$ multiple M2-branes on ${\mathbb C}^4/{\mathbb Z}_k$ is described by the 
 ${\mathcal N}=6$ Chern-Simons-matter theory
which celebrates the gauge group $U(N)\times U(N)$ and levels $k$ and $-k$.
Before this important discovery, the search of
worldvolume theory of multiple M2-branes by supersymmetrizing the
three-dimensional Chern-Simons theory begins to the pioneering
studies in \cite{Sa}. 
The study of three dimensional conformal field theories is relevant   in condensed matter systems  also as they
could describe interesting conformal fixed points.  More recently, the study on perturbative part in the ABJM theory resulting a novel instanton contribution in the orbifold theory has also been made in \cite{hami}. The ABJM theory follows less supersymmetries than
the three-dimensional Chern-Simons theory   constructed by Bagger, Lambert and  Gustavsson \cite{blg,blg1,blg2,blg3,blg4} which follows   ${\mathcal N}=8$ supersymmetry. The BLG theory was  
 conjectured to be related to a specific theory of M2-branes for k = 1, 2 \cite{mukhi,lam}.  The ${\mathcal N}=1$ supersymmetric
higher-order terms that follow from the BLG theory in its expansion with respect to the inverse squared  gauge coupling constant is analysed in \cite{keto}. The BLG theory follows a Lie three-algebra. The underlying gauge symmetry of the theory is an ordinary gauge theory based on Lie algebras \cite{lamb}.

As it is well known, whatever the scheme employed to quantize a gauge theory, a gauge-fixing 
is required in order to keep on the quantization program. The gauge-fixing  can be implemented to the theory  by adding a non-invariant term,  so called 
gauge-breaking term, to the classical action. Consequently, the resulting (effective)
 action becomes a gauge parameter dependent functional on the field configuration manifold.
However, the gauge invariance of the quantum
theory is desired because the expectation values of physical quantities become
independent of the choice of a gauge condition. The best way to realize the gauge independence is to observe  the on-shell   quantum effective action,  evaluated at those configurations that extremize it,  when estimating S-matrix elements (or expectation values) of 
the gauge independent quantities. 
 According to the Nielsen identities \cite{nil}, the variation of the quantum effective
action due to changes in the functions that fix the gauge is linear in the quantum corrected
equations of motion for the mean fields which follows that the on-shell quantum
effective action does not depend on the choice of the gauge breaking term. Even though the mean fields do depend on the gauge-fixing, but this dependence gets canceled by the explicit gauge-fixing dependence of the quantum effective action \cite{del,del1,del2}.

Apart from such investigations, the BRST quantization \cite{sor} of both the BLG and the  ABJM theory have been explored
in recent past \cite{sud0,sud1,sud2,sud20,sud24, sudhak1}. For instance,  the spontaneous breaking of BRST symmetry in ABJM theory in both non-linear as well as in maximally Abelian gauge and the occurrence of the ghost condensation have been studied \cite{sud1}.
The connection of different gauges through generalized BRST transformation 
is investigated in \cite{sud2}. 
However the Batalin-Vilkovisky quatization of ABJM theory is analysed in \cite{sudd}.
The Ward identities and gauge flow for M-theory in ${\cal N}{=}3$ superspace has been 
studied recently \cite{sudhak} where   the gauge dependence of one-particle irreducible amplitudes in such superconformal Chern-Simons theory is shown to be generated by a canonical flow with respect to the extended Slavnov-Taylor identity.
   The ${\cal N}=2$ supersymmetric Chern-Simons theory coupled to matter fields is studied 
   in the large $N$ limit and the two-loop anomalous  dimensions of certain operators
   are also computed \cite{xi}.
Although these progresses have been made towards the complete understanding of the ABJM 
theory, Ward identities as well as
the Nielsen identities for the two-point functions of the ABJM theory which 
guarantees the physical observables to be  gauge independent have not been studied yet.  
This provides a motivation to us for the present investigations.

The aim of this paper is to investigate explicitly the Ward identities at quantum level to show the renormalizability of the model algebraically and  to compute Nielsen
identities for $\mathcal{N} =1$ ABJM theory which will be helpful in the investigations of the gauge dependence of the effective potential. First of all, we 
review the  ABJM theory in $\mathcal{N} =1$
superspace with their  gauge structure. The BRST quantization as well as
Ward identities for the 
model are analysed in the covariant gauge. The
Slavnov-Taylor identities, gauge condition, antighost equation, ghost number and
spinor number  are also computed. 
Subsequently, we discuss  the renormalizability of the theory with the help of  gauge conditions 
as well as antighost equation at all order. We show that these ward identities hold at quantum level.
We derive the Nielsen identities Green's function for ABJM theory following the method
discussed by Piguet and Sibold in \cite{pigu}.
 It is evident in what follows that the Nielsen identities are  very helpful in investigations of on-mass shell Green's functions and on-shell renormalization constants.
 So, the present investigations may be helpful to the investigations of the gauge dependence of the effective potential in    $\mathcal{N} =1$ ABJM theory as well as in the
gauge independence of the physical poles of the propagator.

The rest of the   paper are assembled as following. In section II, we present the 
preliminaries of the
 the ABJM theory in $\mathcal{N} =1$
superspace  and show how it leads to a gauge symmetry. In section III, we 
quantize the model utilizing Faddeev-Popov method and compute the BRST symmetry
and consequently Slavnov-Taylor identities of the effective action. In section IV, we
derive the Nielsen identities for the two-point functions of  $\mathcal{N} =1$
the ABJM theory which gives relation between various Green's functions.
\section{The   ABJM Theory in ${\cal N}=1$ Superspace}
In this section, we recapitulate  the Lagrangian construction of     $\mathcal{N} =1$ ABJM model \cite{ket, mi}. The 
generators of $\mathcal{N} =1$ supersymmetry is given by $Q_a = \partial_a -(\gamma^\mu \partial_\mu)^b_a \theta_b$, where  $\theta^a$ is a two component anti-commutating parameter used to specify the three dimensional $\mathcal{N} =1$ superspace together with the three spacetime coordinates.  To describe the ABJM theory in ${\cal N}=1$ superspace,
we first define the Chern-Simons Lagrangian   $\mathcal{L}_{CS} $ as follows 
 \begin{eqnarray}
 \mathcal{L}_{CS} &=& \frac{k}{2\pi} \int d^2 \,  \theta \, \, 
  Tr \left[  \Gamma^a          \Upsilon_a - \tilde{\Gamma}^a         \tilde{\Upsilon}_a
\right], 
\end{eqnarray}
where $k$ is an integer  and 
 \begin{eqnarray}
\Upsilon_a & = &  \frac{1}{2} D^b D_a \Gamma_b - \frac{i}{2} 
 [\Gamma^b , D_b \Gamma_a]    -
 \frac{1}{6} [ \Gamma^b ,
\{ \Gamma_b , \Gamma_a\}    ]  - \frac{1}{6}[\Gamma^b, \Gamma_{ab}],    \\
 \Gamma_{ab} & = & -\frac{i}{2} [ D_{(a}\Gamma_{b)} 
- i\{\Gamma_a, \Gamma_b\}    ],\nonumber \\
\tilde \Upsilon_a & = &\frac{1}{2} D^b D_a \tilde \Gamma_b 
- \frac{i}{2}  [\tilde \Gamma^b , D_b \tilde\Gamma_a]    -
 \frac{1}{6} [ \tilde \Gamma^b ,
\{ \tilde \Gamma_b ,  \tilde \Gamma_a\} ]  - \frac{1}{6}
[\tilde \Gamma^b, \tilde \Gamma_{ab}],    \\
 \tilde \Gamma_{ab} & = & -\frac{i}{2} [ D_{(a}\tilde \Gamma_{b)} 
- i\{\tilde \Gamma_a, \tilde \Gamma_b\}    ].
\end{eqnarray} 
Here the gauge  superfields  $\Gamma_a$ and $\tilde \Gamma_a$
are   matrix valued spinor superfields suitably contracted with generator $T_A$
of   Lie algebra as follows: $\Gamma_a= \Gamma_a^A T_A$ and $\tilde\Gamma_a= \tilde\Gamma_a^A T_A$, 
respectively. The generators  $Q_a$ commute  with the superspace derivative,  
$D_a = \partial_a + (\gamma^\mu \partial_\mu)^b_a \theta_b$, which plays an important role in   construction of the 
Lagrangian    for ABJM theory in $\mathcal{N} =1$
superspace.
In component form,  these superfields are expressed  by
\begin{eqnarray}
 \Gamma_a = \chi_a + B \theta_a + \frac{1}{2}(\gamma^\mu)_a A_\mu + i\theta^2 \left[\lambda_a -
 \frac{1}{2}(\gamma^\mu \partial_\mu \chi)_a\right], \nonumber \\
 \tilde\Gamma_a = \tilde\chi_a + \tilde B \theta_a + \frac{1}{2}(\gamma^\mu)_a \tilde A_\mu + i\theta^2 \left[\tilde \lambda_a -
 \frac{1}{2}(\gamma^\mu \partial_\mu \tilde\chi)_a\right]. 
 \end{eqnarray}
 The Lagrangian    for the matter sector $\mathcal{L}_{M}$ is given by 
\begin{eqnarray}
 \mathcal{L}_{M} &=& \frac{1}{4} \int d^2 \,  \theta \, \,  
Tr \left[  \nabla^a_{}         X^{I \dagger}               
\nabla_{a 
}               X_I     + 
  \mathcal{V}_{        } \right],
\end{eqnarray}
with  the super-covariant derivatives  of matrix valued complex scalar superfields
$ X^I$ and $X^{I  \dagger}$,
\begin{eqnarray}
 \nabla_{a}              X^{I } &=& D_a  X^{I } + i \Gamma_a          
    X^I - i  X^I        \tilde\Gamma_a      , \nonumber \\ 
 \nabla_{a}              X^{I \dagger} &=& D_a  X^{I  \dagger} 
- i X^{I  \dagger}       \Gamma_a    
        + i \tilde\Gamma_a            X^{I  \dagger}, 
\end{eqnarray}
and the potential term  $\mathcal{V}$,
\begin{eqnarray}
 \mathcal{V}      \propto
[ X_I       X^{I \dagger}        X_J       X^{J\dagger}  X_K       X^{K\dagger}]. 
\end{eqnarray}
With the help of  Chern-Simons and matter Lagrangian, 
the  Lagrangian   for ABJM theory having  gauge group $U(N)_{k}  \times U(N)_{-k}$  is 
written by
\begin{equation}
{ \mathcal{L}_c} =  \mathcal{L}_{M} + \mathcal{L}_{CS}.\label{cl}
\end{equation}  
The gauge symmetry of ABJM follows $U(N)\times U(N)$ group. 
As the Lagrangian for each gauge superfield consists
  a Chern-Simons term,  the gauge invariance requires the coupling constant 
to be integer valued \cite{jac,jac1}.
Under a gauge transformation the scalars and
gauge superfields transform as 
\begin{eqnarray}
 \delta \Gamma_a =  \nabla_a {     } \Lambda, 
&&   \delta \tilde\Gamma_a = \tilde\nabla_a {     }
 \tilde\Lambda, \nonumber \\ 
\delta X^{I } = i(\Lambda{     } X^{I }  - X^{I }{     }\tilde \Lambda ),  
&&  \delta  X^{I \dagger  } 
= i(   \tilde \Lambda {     } X^{I\dagger  }-X^{I\dagger  }{     } \Lambda), 
\end{eqnarray}
where $ \Lambda = \Lambda^A T_A$ and $\tilde \Lambda = \tilde \Lambda^A\tilde  T_A$ are the 
global transformation parameters.  The above transformations leave the classical Lagrangian of the model (\ref{cl}) invariant.
\section{ ABJM Theory: Slavnov-Taylor identity}
In order to give quantum description the ABJM theory,  one must add the gauge-fixing term and the corresponding Faddeev-Popov term  to the invariant Lagrangian (\ref{cl})  \cite{mir,sudd}. By doing so, the gauge fixing  term breaks the gauge invariance and, thus, removes the divergence of the functional integral. However, the  Faddeev-Popov term improves the integration measure to provide correct predictions for gauge invariant observables. 
 Here, the gauge superfields satisfy  the following   gauge  conditions: $
G_1 \equiv D^a \Gamma_a  =0,\,  \tilde G_1 \equiv D^a \tilde{\Gamma}_a =0
$.
The corresponding    gauge-fixing term with gauge parameter $\alpha$  is constructed by
\begin{equation}
\mathcal{L}_{gf} = \int d^2 \,  \theta \, \,  \mbox{Tr}   \left[ b   (D^a \Gamma_a) + \frac{\alpha}{2}b   b -
  \tilde{b}    (D^a \tilde{\Gamma}_a) - \frac{\alpha}{2}\tilde{b}    \tilde b 
\right],
\end{equation}
where $b$ and $\tilde b$ are Nakanishi-Lautrup type auxiliary fields.
With the help of of ghost fields $c, \tilde{c}$ and corresponding anti-ghost fields $\bar c,  \tilde{\bar{c}}$, the Faddeev-Popov ghost term is written explicitly
by  
\begin{equation}
\mathcal{L}_{gh} =-\int d^2 \,  \theta \, \,    \mbox{Tr} 
[  \bar{c}    D^a \nabla_a    c - \tilde{\bar{c}}    D^a \tilde{\nabla}_a   \tilde{c} ].
\end{equation}
The sum of  gauge fixing and ghost terms  is defined by  
\begin{equation}
{\cal L}_{g}={\cal L}_{gf} + {\cal L}_{gh},
  \end{equation} which is BRST exact and hence justifies its own  BRST invariance due to
 the nilpotency property. 
For the present ABJM model, the nilpotent  BRST transformations (i.e. $\delta_b^2=0$) are 
\begin{eqnarray}
&&\delta_b \,\Gamma_{a} = \lambda\nabla_a    c,\ \   \delta_b \, \tilde\Gamma_{a} = \tilde\lambda\tilde\nabla_a    
 \tilde c, \nonumber \\
 &&\delta_b  \,c = -  \frac{1}{2} \lambda {[c,c]}_ { }, \ \delta_b  \,\tilde{ {c}} = -  \frac{1}{2}\tilde \lambda [\tilde{ {c}} ,  \tilde c]_{ }, \nonumber \\
 &&\delta_b  \,\bar{c} =  \lambda b,\ \ \ \ \  \delta_b  \,\tilde {\bar c} = \tilde \lambda \tilde b, \nonumber \\ 
&&\delta_b  \,b =0,\ \ \ \ \ \ \delta_b  \, \tilde b= 0, \nonumber \\ 
 &&\delta_b  \, X^{I } = i \lambda c   X^{I } -  iX^{I }  \tilde c\tilde  \lambda,\nonumber \\ 
 && 
  \delta_b  \, X^{I \dagger }
 = i   \tilde  \lambda\tilde c    X^{I \dagger } - i  X^{I \dagger }  c \lambda,\label{brs}
\end{eqnarray}
where $\lambda$ and $\tilde\lambda$ are Grassmannian parameters.
The  effective ABJM Lagrangian,  defined by the sum of classical and BRST-exact parts ($ {\cal L}_{ABJM}={\cal L}_c + {\cal L}_{g} 
  $),  is invariant under the above BRST transformations.
 In terms of gauge-fixing fermion, the gauge-fixing and ghost parts of the 
  effective Lagrangian  can also be expressed as
  \begin{eqnarray}
  {\cal L}_g =is_b\int d^2 \,  \theta \, \,    \mbox{Tr} \left[  \tilde{\bar c}  D^a \tilde\Gamma_a +\frac{\alpha}{2}\tilde{b} - \bar c  D^a \Gamma_a -  \frac{\alpha}{2} {b} 
 \right].
  \end{eqnarray}
  Since,   the original action is modified by a BRST-exact piece only, which cannot alter the BRST cohomology, and thereby, in turn, cannot alter the notion of physical states.
  
The Ward identities are   obtained by exploiting this invariance by adding   sources 
corresponding to the non-linear transformations.
For this,  we first   couple a source to each non-linear variation of the fields, i.e. to $s_b\Gamma_a, s_b\tilde{\Gamma}_a, s_bc, s_b\tilde{c}, s_bX^I$ and $s_bX^{I\dag}$. Then, we add   them together to write the auxiliary part as following:
\begin{eqnarray}
{\cal L}_{ext} &=&\int d^2\,  \theta \, \,    \mbox{Tr} \left[ K^a(\nabla_a c) -\tilde K^a(\tilde\nabla_a \tilde c)
-\frac{1}{2}\bar K_c [c,c]+ \frac{1}{2}\tilde{\bar K}_c [\tilde c, \tilde c]
\right.\nonumber\\
&+&\left.  \bar K_{I}(i c   X^{I } -  iX^{I }  \tilde c)+( i   \tilde c    X^{I \dagger } - i  X^{I \dagger }  c)  K_{ I}  \right],
\end{eqnarray}
whereby $K^a,\tilde K^a$
are the Grassmann supersources, $\bar K_c, \tilde{\bar K}_c$ are the bosonic supersources and
$\bar K_{I}$ and $ K_{ I}$  are the matrix valued supersources.
Now, the effective action    
\begin{equation}
\Sigma =\int d^3x \left({\cal L}_c+{\cal L}_{g}  + {\cal L}_{ext}\right),
\end{equation}
leads to the following identities.
\begin{itemize}
\item The Slavnov-Taylor identity is   given by
\begin{eqnarray}
{\cal S} (\Sigma) &=&\int d^3xd^2\theta \left[ \frac{\delta\Sigma}{\delta K^a}\frac{\delta\Sigma}{\delta \Gamma_a}+\frac{\delta\Sigma}{\delta \tilde K^a}\frac{\delta\Sigma}{\delta \tilde\Gamma_a} +\frac{\delta\Sigma}{\delta \bar K_c}\frac{\delta\Sigma}{\delta c}+\frac{\delta\Sigma}{\delta \tilde{\bar K}_c}\frac{\delta\Sigma}{\delta \tilde c}+b\frac{\delta\Sigma}{\delta \bar c}+\tilde b\frac{\delta\Sigma}{\delta \tilde{\bar c}}\right.\nonumber\\
&+&\left. \frac{\delta\Sigma}{\delta \bar K_I}\frac{\delta\Sigma}{\delta X^I} -\frac{\delta\Sigma}{\delta  K_I}\frac{\delta\Sigma}{\delta X^{I\dag}}\right]=0. 
\end{eqnarray}
\item The gauge conditions are given by
\begin{eqnarray}
\frac{\delta\Sigma}{\delta b}&=& D^a\Gamma_a +\alpha b,\nonumber\\
\frac{\delta\Sigma}{\delta \tilde b}&=& D^a\tilde \Gamma_a +\alpha \tilde b,\label{gauge}
\end{eqnarray}
Though these symmetries are linearly broken,  these are allowed due to the
Quantum Action Principle (QAP) \cite{jh,ym,te,sor}.
\item The   action enjoys anti-ghost equations  
\begin{eqnarray}
&&\left(\frac{\delta}{\delta \bar c} -D_a\frac{\delta}{\delta K_a}\right)\Sigma =0,\nonumber\\
&&\left(\frac{\delta}{\delta \tilde{\bar c}} -D_a\frac{\delta}{\delta \tilde K_a}\right)\Sigma =0.\label{an}
\end{eqnarray}
\item
We also notice that the action preserves the ghost
number  
\begin{eqnarray}
 \mathcal{G} n(\Sigma)&=&\int d^3xd^2\theta \left[c\frac{\delta}{\delta c} -\tilde c\frac{\delta}{\delta \tilde c} -\bar c\frac{\delta}{\delta \bar c}-\tilde{\bar c}\frac{\delta}{\delta \tilde{\bar c}}- K^a\frac{\delta}{\delta K^a}- \tilde K^a\frac{\delta}{\delta \tilde K^a} - 2 \bar K_c^a\frac{\delta}{\delta \bar K_c^a}\right.\nonumber\\
 &-& \left.  2 \tilde{\bar K}_c^a\frac{\delta}{\delta \tilde{\bar K}_c^a}-  \bar K_I\frac{\delta}{\delta \bar K_I}-    K_I\frac{\delta}{\delta   K_I}\right]\Sigma =0.
\end{eqnarray}
\item The action also preserves  the spinor number
\begin{eqnarray}
 \mathcal{S} n(\Sigma)&=&\int d^3xd^2\theta \left[ X_I\frac{\delta}{\delta X_I}-
 X^\dag_I\frac{\delta}{\delta X^\dag_I}+K_I\frac{\delta}{\delta K_I}-
 \bar K_I\frac{\delta}{\delta\bar K_I}\right]\Sigma =0.
\end{eqnarray}

\end{itemize}
These identities will be very helpful to show the algebraic renormalizability
of the ${\cal N}=1$ ABJM theory. 
However, we should notice that the spinor number is not a necessary Ward identity to
prove the renormalizability. The counter terms can also be derived by 
following these identities.

\section{Ward identities at the quantum level}
Now, by considering gauge conditions and antighost equations, we try to prove that all the Ward identities can be transformed to the quantum
level.
\subsubsection{The gauge conditions}
Let us start with the gauge conditions (\ref{gauge}), we would like to prove that  
identities (\ref{gauge}) are not affected by the radiative corrections. 

The QAP \cite{sor} translates the symmetry to the quantum level as 
\begin{eqnarray}
\frac{\delta\Sigma}{\delta b}&=& D^a\Gamma_a +\alpha b+  \bar{\Delta}\cdot \Sigma,\nonumber\\
\frac{\delta\Sigma}{\delta \tilde b}&=& D^a\tilde \Gamma_a +\alpha \tilde b+  
\tilde{\bar{\Delta}}\cdot \Sigma. 
\end{eqnarray}
As we know that  $\bar{\Delta}$ and $\tilde{\bar{\Delta}}$ can only start from order $\hbar$, so let us  assume that these start  with order  $\hbar^n$, $n\geq 1$
\begin{eqnarray}
\frac{\delta\Sigma}{\delta b}&=& D^a\Gamma_a +\alpha b+ \hbar^n \bar{\Delta}+ O(\hbar^{n+1}),\nonumber\\
\frac{\delta\Sigma}{\delta \tilde b}&=& D^a\tilde \Gamma_a +\alpha \tilde b+ \hbar^n 
\tilde{\bar{\Delta}}+ O(\hbar^{n+1}), 
\end{eqnarray}
where $\bar{\Delta}$ and $\tilde{\bar{\Delta}}$ are the local polynomials
of the sources and superfields of dimensions $3/2$ with ghost number zero and therefore given by
\begin{eqnarray}
\bar{\Delta}(x) &=& F(\Gamma^a, c, \bar{c})(x) +\omega b(x), \nonumber\\
\tilde{\bar{\Delta}}(x) &=& \tilde F(\tilde \Gamma^a, \tilde c, \tilde {\bar{c}})(x) +\tilde\omega 
\tilde b(x), \label{del1}
\end{eqnarray}
written in terms of the local polynomials $F, \tilde F$ and constants $\omega$ and $\tilde \omega$.
Here, we assume that identities (\ref{gauge})   hold
below the order $n$ in $\hbar$  and, therefore, the most general breaking are compatible with the power-counting constraints above. Now, the consistency conditions 
\begin{eqnarray}
\frac{\delta}{\delta b(x)}\bar \Delta (y)- \frac{\delta}{\delta b(y)}\bar \Delta(x) =0,\ \ 
\frac{\delta}{\delta \tilde b(x)}{\tilde {\bar \Delta}} (y)- \frac{\delta}{\delta \tilde b(y)}\tilde{\bar \Delta}(x) =0, \label{con}
\end{eqnarray}
follow from the facts that $[\delta/\delta b(x), \delta/\delta b(y)] =0$,   $[\delta/\delta \tilde b(x), \delta/\delta \tilde b(y)] =0$, respectively. Utilizing 
 (\ref{del1}) and (\ref{con}), it is easy to write
\begin{eqnarray}
\bar{\Delta} (x)&=& \frac{\delta}{\delta b(x)}\int d^3 y\left[F(\Gamma^a, c, \bar{c})(y) +
\frac{1}{2}\omega b(y)b(y)\right], \nonumber\\
\tilde{\bar{\Delta}} (x)&=&  \frac{\delta}{\delta \tilde b(x)}\int d^3 y\left[\tilde F(\tilde \Gamma^a, \tilde c, \tilde {\bar{c}})(y) +\frac{1}{2}\tilde \omega 
  \tilde b(y)\tilde b(y)\right].
\end{eqnarray}
Now, we  redefine the effective action as
\begin{eqnarray}
\bar\Sigma =\Sigma -\hbar^n \int d^3 y\left[F(\Gamma^a, c, \bar{c})(y) +
\frac{1}{2}\omega b(y)b(y)\right] +\hbar^n\int d^3 y\left[\tilde F(\tilde \Gamma^a, \tilde c, \tilde {\bar{c}})(y) +\frac{1}{2}\tilde \omega 
  \tilde b(y)\tilde b(y)\right],\label{def}
\end{eqnarray}
which follow,
\begin{eqnarray}
\frac{\delta\bar\Sigma}{\delta b}&=& D^a\Gamma_a +\alpha b+   O(\hbar^{n+1}),\nonumber\\
\frac{\delta\bar\Sigma}{\delta \tilde b}&=& D^a\tilde \Gamma_a +\alpha \tilde b+ O(\hbar^{n+1}). 
\end{eqnarray}
We repeat this argument at each consecutive order. Consequently, this ends the recursive proof of the renormalizability of the gauge conditions.
\subsubsection{Antighost equations}
Let us now investigate the antighost equations (\ref{an})  to prove that
\begin{eqnarray}
&&\left(\frac{\delta}{\delta \bar c} -D_a\frac{\delta}{\delta K_a}\right)\bar\Sigma =0,\nonumber\\
&&\left(\frac{\delta}{\delta \tilde{\bar c}} -D_a\frac{\delta}{\delta \tilde K_a}\right)\bar\Sigma =0,
\end{eqnarray}
where $\bar\Sigma$ has already been defined in (\ref{def}). To write these equations into a
more simple  form, we redefine the superfields to yield
\begin{eqnarray}
&&\frac{\delta}{\delta K_a} =\frac{\delta}{\delta \hat K_a},\ \ \ \frac{\delta}{\delta \bar c}= \frac{\delta}{\delta \hat{\bar c}}-D_a\frac{\delta}{\delta \hat K_a},\nonumber\\
&&\frac{\delta}{\delta \tilde K_a} =\frac{\delta}{\delta \hat {\tilde K}_a},\ \ \ \frac{\delta}{\delta \tilde{\bar c}}= \frac{\delta}{\delta \hat{\tilde{\bar c}}}-D_a\frac{\delta}{\delta \hat {\tilde K}_a}.
\end{eqnarray}
Thus, the antighost equations become 
\begin{eqnarray}
\frac{\delta}{\delta \bar c}\hat\Sigma =0,\ \ \ \
\frac{\delta}{\delta \tilde{\bar c}}\hat\Sigma =0,
\end{eqnarray}
where $\hat\Sigma$ is the effective action written for new variables  ($\hat K_a, \hat {\tilde K}_a, \hat{\bar c}, \hat{\tilde{\bar c}}$). Now, we apply QAP and thus obtain
\begin{eqnarray}
\frac{\delta}{\delta\hat{\bar c}}\hat \Sigma=\bar\Delta \hat{\Sigma},\ \ \ \ \frac{\delta}{\delta\hat{\tilde{\bar c}}}\hat \Sigma=\tilde {\bar \Delta} \hat{\Sigma}.
\end{eqnarray}
Here, we assume again that the breaking  starts at order $\hbar^n$, with  $n\geq 1$,
\begin{eqnarray}
\frac{\delta}{\delta\hat{\bar c}}\hat \Sigma=\hbar^n\bar\Delta +   O(\hbar^{n+1}),\ \ \ \ \frac{\delta}{\delta\hat{\tilde{\bar c}}}\hat \Sigma=\hbar^n\tilde {\bar\Delta }+   O(\hbar^{n+1}),
\end{eqnarray}
with  local polynomials of the sources and  superfields  
of dimensions $3/2$ with ghost number  3/2, $\bar\Delta$ and $\tilde{\bar\Delta}$,
\begin{eqnarray}
&&\bar\Delta (x) = G(\Gamma^a, c) (x) +v(c)\hat{\bar c}(x),\nonumber\\
&&\tilde{\bar\Delta} (x) = \tilde G(\tilde \Gamma^a, \tilde c) (x) +\tilde v(\tilde c)\hat{\tilde{\bar c}}(x).
\end{eqnarray}
Here, we found that
\begin{eqnarray}
\frac{\delta}{\delta \hat{\bar c}(x)}\bar \Delta (y)- \frac{\delta}{\delta \hat{\bar c}(y)}\bar \Delta(x) =0,\ \ 
\frac{\delta}{\delta  \hat{\tilde{\bar c}}(x)}{\tilde {\bar \Delta}} (y)- \frac{\delta}{\delta   \hat{\tilde{\bar c}}(y)}\tilde{\bar \Delta}(x) =0,
\end{eqnarray}
which have the following solutions:
\begin{eqnarray}
&&\bar \Delta (x)=\frac{\delta}{\delta \hat{\bar c}(x)}\int d^3y\left( \hat{\bar c}G(\Gamma^a, c) (y) +\frac{1}{2}v(c)\hat{\bar c} \hat{\bar c}(y) \right),\nonumber\\
&&\tilde{\bar \Delta} (x) =\frac{\delta}{\delta \hat{\tilde{\bar c}}(x)}\int d^3y\left( \hat{\tilde{\bar c}}\tilde G(\tilde\Gamma^a, \tilde c) (y) +\frac{1}{2}\tilde v(\tilde c)\hat{\tilde{\bar c}} \hat{\tilde{\bar c}}(y) \right),
\end{eqnarray}
We can redefine the action analogously as in equation (\ref{def}), so the antighosts as well as the
gauge conditions hold to order $\hbar^n$.  Similarly, we are able to prove that all  these identities hold to all orders.
\section{Nielsen identity for ABJM theory}
In this section, we analyse the Nielsen identity for the ABJM theory in ${\cal N}=1$
superspace following \cite{pigu,bre}. To do so, we first perform a shift in the Lagrangian density as follows:
\begin{eqnarray}
{\cal L}_{ABJM} \rightarrow{\cal L}'_{ABJM} ={\cal L}_{ABJM} +\int d^2 \,  \theta \, \,    \mbox{Tr} \left(\frac{\chi}{2}\bar c b -\frac{\chi}{2}\tilde{ \bar c} \tilde b\right),\label{bh}
\end{eqnarray}
where $\chi$  is a    global Grassmannian variables, i.e., $\chi^2=0$.
It is clear upon a little reflection that  
  this extra term  does not change the dynamics of the
theory.
The resulting Lagrangian (\ref{bh}) remains invariant under the following
extended set of BRST transformations:
\begin{eqnarray}
 &&\delta^+_b \,\Gamma_{a} =  \lambda\nabla_a    c,\ \delta^+_b \, \tilde\Gamma_{a} =\tilde  \lambda\tilde\nabla_a    
 \tilde c, \nonumber \\
 &&\delta^+_b  \,c = -  \frac{1}{2} \lambda {[c,c]}_ { },\  \delta^+_b  \,\tilde{ {c}} = -  \frac{1}{2}\tilde  \lambda [\tilde{ {c}} ,  \tilde c]_{ }, \nonumber \\
 && \delta^+_b  \,\bar{c} =  \lambda b,\ \ \ \delta^+_b  \,\tilde {\bar c} = \tilde \lambda \tilde b, \nonumber \\ 
&&\delta^+_b  \,b =0,\ \ \ \ \  \delta^+_b  \, \tilde b= 0, \nonumber \\ 
 &&\delta^+_b  \,\alpha = \lambda\chi,\ \ \  \ \ \delta^+_b  \, \chi= 0, \nonumber \\ 
 && \delta^+_b  \, X^{I } = i \lambda c   X^{I } -  iX^{I }  \tilde c\tilde  \lambda,  \nonumber \\ 
 && 
 \delta^+_b  \, X^{I \dagger }
 = i \tilde \lambda  \tilde c    X^{I \dagger } - i  X^{I \dagger }  c \lambda,\label{brs1}
  \end{eqnarray}
  where $\lambda$ and $\tilde{\lambda}$ are the Grassmann parameters. 
The interesting  point noted here is that the gauge parameter also changes under the
transformation.
To exploit this invariance to derive the Nielsen
identities, we   construct the following generating functional:
\begin{eqnarray}
Z &=&\int [{\cal D}\phi]\exp\left[i\int d^3x \left({\cal L}'_{ABJM} +\int d^2 \,  \theta \, \,    \mbox{Tr} \left\lbrace J^a\Gamma_a -  \tilde J^a\tilde\Gamma_a  
+\bar J^{I }X_I+X_I^{\dag}J^I  + bJ_b -\tilde{b}J_{\tilde b}
\right.\right.\right.\nonumber\\
&+&\left.\left.\left. \bar J_c c+
\bar c J_c - \tilde{\bar J}_c \tilde c-
\tilde{\bar c} \tilde J_c +K^a(\nabla_a c) -\tilde K^a(\tilde\nabla_a \tilde c)
-\frac{1}{2}\bar K_c [c,c]+ \frac{1}{2}\tilde{\bar K}_c [\tilde c, \tilde c]
\right.\right.\right.\nonumber\\
&+&\left.\left.\left. \bar K_{I}(i c   X^{I } -  iX^{I }  \tilde c)+( i   \tilde c    X^{I \dagger } - i  X^{I \dagger }  c)  K_{ I}\right\rbrace
\right) \right].
\end{eqnarray}
The various sources denoted by $J$ with a different subscript are
the obvious ones, however, the purpose of the additional, rather exotic looking, sources denoted by $K$'s will become apparent in a moment.
The terms with such additional sources of the Lagrangian
may be rewritten as
\begin{equation}
\bar K_{I} \left(\frac{\delta^+_b  X^{I }}{\delta\lambda}+\frac{\delta^+_b  X^{I }}{\delta\tilde\lambda} \right) +\left(\frac{\delta^+_b   X^{I \dagger }}{\delta\lambda}+\frac{\delta^+_b   X^{I \dagger }}{\delta\tilde\lambda}\right) K_{ I}.\label{k}
\end{equation}
To study the gauge dependence of the gauge and matter propagators, we now introduce the 
generating
functional of proper Green functions
\begin{eqnarray}
 &&\Delta (\Gamma_a,\tilde \Gamma_a, X_I,X_I^\dag, c,\tilde{c}, \bar{c}, \tilde{\bar{c}}, c, \tilde{b}, \alpha, \chi, K_a, \tilde K_a, \bar K_{I},  K_{I}) = W(J^a, \tilde J^a, \bar J^{I },J^I,J_b,      J_{\tilde b}, \bar J_c, J_c, \tilde{\bar J}_c,\nonumber\\
 &&  \tilde J_c,K^a, \tilde K^a,
 \bar K_c, \tilde{\bar K}_c, \alpha, \chi, \bar K_{I},  K_{I})  
  - \int d^3x \int d^2 \,  \theta \ \mbox{Tr}[  J^a\Gamma_a -  \tilde J^a\tilde\Gamma_a  
+\bar J^{I }X_I  + X_I^{\dag}J^I+ bJ_b\nonumber\\
&& -\tilde{b}J_{\tilde b}+\bar J_c c+
\bar c J_c - \tilde{\bar J}_c \tilde c  
 - 
\tilde{\bar c} \tilde J_c ].\label{del}
\end{eqnarray}
The invariance of above functional under  (\ref{brs1}) leads to
\begin{eqnarray}
\delta^+_b \Delta \equiv 0&=&\delta^+_b \Gamma_a \frac{\delta\Delta}{\delta\Gamma_a}+
\delta^+_b \tilde\Gamma_a \frac{\delta\Delta}{\delta\tilde\Gamma_a}+
\delta^+_b \bar c \frac{\delta\Delta}{\delta\bar c}+
\delta^+_b \tilde{\bar c} \frac{\delta\Delta}{\delta\tilde{\bar c}}+
\delta^+_b  c \frac{\delta\Delta}{\delta  c}\nonumber\\
&+&
\delta^+_b \tilde c \frac{\delta\Delta}{\delta\tilde c}+
\delta^+_b \alpha \frac{\delta\Delta}{\delta\alpha}+
\delta^+_b X^I \frac{\delta\Delta}{\delta X^I}+
\delta^+_b X^{I\dag} \frac{\delta\Delta}{\delta X^{I\dag}}.\label{xy}
\end{eqnarray}
Here, the terms corresponding to the fields which vanish under the transformation  (\ref{brs1}) 
would not be appear. 
Utilizing (\ref{k}) together with (\ref{del}), we  rewrite (\ref{xy}) as follows: 
\begin{eqnarray}
&&   \frac{\delta\Delta}{\delta K^a}\frac{\delta\Delta}{\delta\Gamma_a}-
\frac{\delta\Delta}{\delta \tilde K^a}   \frac{\delta\Delta}{\delta\tilde\Gamma_a}+
b \frac{\delta\Delta}{\delta\bar c}+
  \tilde b \frac{\delta\Delta}{\delta\tilde{\bar c}}+
\frac{\delta\Delta}{\delta \bar K_c} \frac{\delta\Delta}{\delta  c}\nonumber\\
&&-
 \frac{\delta\Delta}{\delta \tilde{\bar K}_c } \frac{\delta\Delta}{\delta\tilde c}+
\chi \frac{\delta\Delta}{\delta\alpha}+
\frac{\delta\Delta}{\delta \bar K_I} \frac{\delta\Delta}{\delta X^I}+
\frac{\delta\Delta}{\delta K_I} \frac{\delta\Delta}{\delta X^{I\dag}} =0.\label{nel}
\end{eqnarray}
Now, we are able to have the Nielsen identities
for the ${\cal N}=1$ ABJM theory.
For this, we differentiate equation (\ref{nel}) with respect to $\chi$ and then set $\chi=0$.
This yields
\begin{eqnarray}
&&   \frac{\delta\Delta}{\delta\alpha}+  \frac{\delta^2\Delta}{\delta\chi\delta K^a}\frac{\delta\Delta}{\delta\Gamma_a}-\frac{\delta\Delta}{\delta K^a}
\frac{\delta^2\Delta}{\delta\chi\delta\Gamma_a}    -
\frac{\delta^2\Delta}{\delta\chi\delta \tilde K^a}   \frac{\delta\Delta}{\delta\tilde\Gamma_a}+\frac{\delta\Delta}{\delta \tilde K^a}   \frac{\delta^2\Delta}{\delta\chi\delta\tilde\Gamma_a} +
b \frac{\delta^2\Delta}{\delta\chi\delta\bar c}+
  \tilde b \frac{\delta^2\Delta}{\delta\chi\delta\tilde{\bar c}}\nonumber\\
&&+
\frac{\delta^2\Delta}{\delta\chi\delta \bar K_c} \frac{\delta\Delta}{\delta  c}+
\frac{\delta\Delta}{\delta \bar K_c} \frac{\delta^2\Delta}{\delta\chi\delta  c}-
 \frac{\delta^2\Delta}{\delta\chi\delta \tilde{\bar K}_c } \frac{\delta\Delta}{\delta\tilde c}
 -
 \frac{\delta\Delta}{\delta \tilde{\bar K}_c } \frac{\delta^2\Delta}{\delta\chi\delta\tilde c}+
\frac{\delta^2\Delta}{\delta\chi\delta \bar K_I} \frac{\delta\Delta}{\delta X^I}+
\frac{\delta\Delta}{\delta \bar K_I} \frac{\delta^2\Delta}{\delta\chi\delta X^I}\nonumber\\
&&+
\frac{\delta^2\Delta}{\delta\chi\delta K_I} \frac{\delta\Delta}{\delta X^{I\dag}}+
\frac{\delta\Delta}{\delta K_I} \frac{\delta^2\Delta}{\delta\chi\delta X^{I\dag}}  =0.
\end{eqnarray}
From these results, we can generate the   Nielsen identities for the two-point
functions of ABJM theory.
The investigations of the gauge dependence of the effective potential in the ABJM
theory as well as   the gauge independence of the physical poles of the propagator
can schematically be derived from the above Nielsen identities Green's function. 
\section{Conclusion}
Since M2-branes are three-dimensional objects embedded in an eleven-dimensional
manifold, so the world-volume theory  of such branes
must be a three-dimensional gauge theory. However, in the low-energy limit, the theory
must flow to a non-trivial fixed point. The promising candidate for 
the theories fulfilling  all these requirements  was constructed by Aharony,
Bergman, Jafferis, and Maldacena (ABJM). The ABJM model is a three-dimensional superconformal Chern-Simons-matter theory with gauge group $U(N)\times U(N)$.

In this paper, we have reviewed the gauge symmetry of ABJM theory in ${\cal N}=1$ superspace.
According to the standard quantization method, a theory having gauge symmetry
can be quantizing only after breaking the gauge invariance by   adding a gauge variant
term which induces a ghost term to the action. The resulting action
follows the BRST symmetry. With the help of BRST symmetry, we have computed the Slavnov-Taylor
identities, gauge condition, anti-ghost equation, ghost number and spinor number. With the help of   Ward identities, namely,  gauge condition and anti-ghost equation at quantum level, 
we established the renormalizability  of the ABJM theory in ${\cal N}=1$ superspace
at all order.
Further, we have investigated the Nielsen identities for the
two-point functions of ABJM theory in ${\cal N}=1$ superspace in the covariant
formalism.  The
Nielsen identities offer possibilities to check one's calculations,
however, they also allow us to see where physical
meaning may be found in apparently gauge dependent Green's functions.  
The present investigations will be very helpful to show the gauge dependence
of the effective potential in a gauge theory (with scalar fields) as
well as the gauge independence of the physical poles of the propagator and on-shell renormalization constants.
Since the on-shell renormalization scheme is not commonly used in ABJM theory, so it will be subject of future investigation.

\end{document}